\documentclass{article}

\usepackage[english]{babel}
\usepackage{amssymb,amsmath,graphicx,algorithm,algpseudocode,framed}
\usepackage[colorlinks=true, allcolors=blue]{hyperref}
\usepackage[letterpaper,top=2cm,bottom=2cm,left=3cm,right=3cm,marginparwidth=1.75cm]{geometry}
\usepackage{indentfirst}
\usepackage{bm}
\usepackage{subcaption}
\usepackage[percent]{overpic}

\usepackage[
  backend=biber,
  style=phys, 
  giveninits=true,
  maxnames=99,
  minnames=99,
  sorting=none,
  doi=false,
  url=false,
  isbn=false,
  eprint=true
]{biblatex}

\AtEveryBibitem{%
  \clearfield{abstract}%
  \clearfield{note}%
  \clearfield{file}%
  \clearfield{urldate}%
  \clearfield{month}
  \clearfield{day}
  \clearlist{language}%
}

\DeclareFieldFormat[article]{volume}{\mkbibbold{#1}}
\DeclareFieldFormat[article]{number}{\mkbibparens{#1}}
\DeclareFieldFormat[misc]{title}{\mkbibquote{#1}}

\DeclareBibliographyDriver{misc}{%
  \usebibmacro{bibindex}%
  \usebibmacro{begentry}%
  \usebibmacro{author/editor+others/translator+others}%
  \setunit{\printdelim{nametitledelim}}\newblock
  \usebibmacro{title}%
  \newunit\newblock
  \usebibmacro{doi+eprint+url}%
  \setunit{\addspace}%
  \printtext[parens]{\usebibmacro{date}}%
  \newunit\newblock
  \usebibmacro{addendum+pubstate}%
  \setunit{\bibpagerefpunct}\newblock
  \usebibmacro{pageref}%
  \usebibmacro{finentry}}

\renewbibmacro*{volume+number+eid}{%
  \printfield{volume}%
  \printfield{number}%
  \printfield{eid}%
}

\newcommand{\bv}{\bm{v}}
\newcommand{\bw}{\bm{w}}
\newcommand{\bu}{\bm{u}}
\newcommand{\rd}{\,\mathrm{d}}
\newcommand{\im}{\mathrm{i}}

\title{
    \vspace{-2em}
    \textbf{\Large A Multi-Species Reactive-Boltzmann Formulation for Self-Consistent Kinetic Simulation of Burning Fusion Plasmas}
}
\date{}

\begin{document}
\author{Mark Dunn\footnote{Aerospace and Energetics Research Program, University of Washington, Seattle, WA 98195, USA (markd406@uw.edu).}, \ 
		Jingwei Hu\footnote{Department of Applied Mathematics, University of Washington, Seattle, WA 98195, USA (hujw@uw.edu).}, \
		Uri Shumlak\footnote{Aerospace and Energetics Research Program, University of Washington, Seattle, WA 98195, USA (shumlak@uw.edu).}}

\maketitle

\section*{Abstract}

Discrepancies between experimental data and radiation-hydrodynamic models of burning plasmas at the National Ignition Facility have been attributed, in part, to possible deviations of reactant ion distributions from Maxwellian equilibrium. In particular, it has been hypothesized that the collisional relaxation of energetic fusion products with the bulk plasma may generate suprathermal ion populations not captured by reduced models. To assess this hypothesis in a fully kinetic setting, we present a multi-species reactive--elastic kinetic framework, in which fusion sources and sinks are evaluated directly from the reactant velocity distribution function with a reactive Boltzmann collision operator. For computational efficiency, elastic interactions are modeled in the small-angle collision limit using the Landau and Lenard--Bernstein collision operators, although the framework admits large-angle elastic scattering generalizations. The integro-differential collision operators are discretized with a fast Fourier spectral method, enabling efficient, high-order accuracy evaluation of reactive processes in three-dimensional velocity space and providing a deterministic alternative to Monte Carlo collision methods. Numerical experiments are performed for spatially homogeneous two- and three-species D--D fusion systems in weakly coupled, early-time regimes. We find that fusion product heating does not generate a significant suprathermal population in the reactant ions: deviation from the corresponding Maxwellian distribution remains small in both the Frobenius norm and the fusion reactivity. 

\section{Introduction}

Thermonuclear fusion devices rely on the deposition of energy from charged fusion products into the background fuel to enter a self-sustained ``burning'' phase and achieve high energy gain. As experimental campaigns approach this regime~\cite{burning_plasma_achieved_NIF,Kiptily2023_JET_burning,SML_FusionSciTech_2024}, accurately capturing the detailed kinetic physics of plasma self-heating (alpha heating in the D--T fusion case) has become increasingly important. Most modeling approaches, including radiation-hydrodynamic~\cite{Marinak2001_HYDRA, Haines_xRAGE_2022, Fryxell_FLASH_2000, Mishchenko2023_numerical_tools_burning_plasmas} and reduced kinetic models~\cite{Reichelt_etal_preprint,Ye_Zhang_Wan_PoP_2025}, assume that fuel ions remain Maxwellian, which greatly simplifies the evaluation of fusion sources and sinks. However, the self-heating process is intrinsically kinetic: fusion products are born approximately monoenergetically in the MeV range, forming a thin spherical shell in velocity space well separated from the thermal population; as they slow down through collisions with the background plasma, their distribution develops a heavy tail extending from thermal energies to the birth energy. It is unknown whether cross-species interactions with these highly non-Maxwellian products create and sustain a significant suprathermal population of fuel ions. However, across both inertial and magnetic confinement fusion, experiments have observed departures from Maxwellian behavior consistent with the presence of suprathermal ion populations~\cite{Hartouni_nature,Salewski2018_ASDEX,Maslov_JET_DT_nonthermal_fusion,Atzeni_burning_plasma_surprise}. Quantifying the regimes in which these effects arise self-consistently is important for predictive modeling, as even modest deviations from Maxwellian equilibrium can significantly impact fusion reactivity and the resulting plasma dynamics.

Previous studies have examined fusion reactivities for prescribed non-Maxwellian reactant distributions, including $\kappa$-distributions \cite{Squarer_Presilla_Onofrio_PRE, Onofrio2018DDreactor} and drift bi-Maxwellian distributions with temperature anisotropy \cite{Xie2023_fusion_reactivities}. These studies show that the computed reactivities can exceed that of a Maxwellian with the same total energy at sufficiently low temperatures due to the strong, non-monotonic dependence of the fusion cross-section on reactant kinetic energy. Related kinetic studies have examined the formation of quasi-steady non-Maxwellian distributions in heated plasmas \cite{Ye_Zhang_Wan_PoP_2025} and the influence of spatial shear-flow gradients on fusion reactivity \cite{Fetsch_Fisch_sheared_flow_reactivity_enhancement}. In the former, the fusion rate remains decoupled from the heating mechanism, which is represented with a fixed source term. In the latter, reactivity is evaluated as a diagnostic, so that fusion reactions do not affect the evolution of the distribution functions. While these works establish that sufficiently strong suprathermal heating can enhance reactivity beyond Maxwellian levels, they do not address whether such distributions can arise and persist self-consistently in a burning plasma, where the heating rate and fusion rate are nonlinearly coupled. This distinction is key to addressing whether such non-Maxwellian features can be sustained without prescribed external sources. 

Self-consistent kinetic simulations of burning D--T plasmas in spatially inhomogeneous spherical geometries have been performed in \cite{vandeWetering2025PICNIC_NES_ICF} using particle-in-cell methods with a Monte Carlo collision routine, which can capture reactions and large- and small-angle elastic scattering efficiently but suffers from statistical noise. The advantage of the deterministic, continuum formulation considered in this work is that it is noise-free with controllable numerical accuracy, enabling the resolution of small deviations from Maxwellian equilibrium in velocity space. In particle-based methods, statistical uncertainty arises from finite sampling and decreases roughly as $\mathcal{O}(N^{-1/2})$ with the number of particles, reflecting convergence in a probabilistic sense. In contrast, continuum discretizations are subject to deterministic discretization error governed by the resolution of velocity space, admitting possibly high-order (e.g., spectral) convergence under grid refinement.

In this work, we present a multi-species continuum kinetic framework in which fusion reactions and collisional relaxation are treated self-consistently without assuming Maxwellian reactant distributions. Fusion sources and sinks are computed directly from the evolving fuel ion velocity distribution functions using a reactive Boltzmann collision operator based on the functional form introduced in \cite{Rossani_Spiga_1999} for chemically reacting gases, in which reactions are modeled as binary collisions that occur only beyond a relative velocity threshold corresponding to the reaction activation energy. While the Boltzmann collision operator provides, in principle, a common framework for elastic and reactive interactions—including rare large-angle Coulomb events and nuclear elastic scattering, we restrict its use here to reactive processes, as the direct numerical evaluation of the Boltzmann--Coulomb operator remains prohibitively expensive for high-dimensional, multi-temporal and velocity-scale continuum kinetic simulations.

Accordingly, we focus here on weakly coupled regimes in which elastic scattering is well-described by reduced models, using either the multi-species Landau operator \cite{Carrillo_Hu_vanFleet_particle_Landau} or a multi-species Lenard--Bernstein (LB) operator \cite{HHHH_2025}. The latter admits an efficient implicit time integration via a moment update procedure, which we extend to systems with both reactive and elastic interactions. Both the reactive Boltzmann and Landau operators are discretized using a fast Fourier spectral method, enabling efficient evaluation in three-dimensional velocity space. For the first time within a continuum kinetic setting, we find that in weakly coupled, spatially homogeneous, early-time D--D fusion regimes, fusion product heating alone does not generate a substantial suprathermal population of fuel ions, and the corresponding fusion reactivity deviates only weakly from its Maxwellian value.

The paper is organized as follows. In Sec.~2, we introduce the multi-species reactive Boltzmann collision operator and its application to effectively irreversible fusion reactions. In Sec.~3, we review the elastic collision models considered in this work, including the multi-species Landau and LB operators. In Sec.~4, we describe the numerical methods, including the fast Fourier spectral method for the reactive Boltzmann and Landau operators and an implicit time-integration scheme for the LB operator. In Sec.~5, we verify the numerical implementation with self-convergence studies and comparison with an analytic solution to the spatially homogeneous multi-species Landau equation. In Sec.~6, we present numerical experiments illustrating the velocity-space dynamics of a burning plasma and provide scalar diagnostics of the deviation from the reactant equivalent Maxwellian distribution. Conclusions and suggestions for future work are given in Sec.~7.

\section{The multi-species reactive Boltzmann collision operator}

Rossani and Spiga \cite{Rossani_Spiga_1999} derived a Boltzmann collision operator for multi-species gas mixtures undergoing reversible chemical reactions with a single internal-energy degree of freedom. Polyatomic generalizations with multiple internal degrees of freedom are available \cite{Borsoni_Bisi_Groppi2022} but are not considered in this work. The framework is easily extended to atomic and nuclear reactions relevant to burning fusion plasmas, such as nuclear fusion, electron-impact ionization, charge exchange, and radiative recombination. 

The use of a Boltzmann-type reactive operator, rather than a simplified Maxwellian-reactivity-based model, enables the characterization of arbitrary deviations from equilibrium and their effect on the velocity-space structure of reactant depletion and product generation. The quantification is first-principles in the sense that the operator relies only on the assumptions of classical Boltzmann kinetic theory (uncorrelated, instantaneous binary interactions). Quantum effects, if present, enter solely through the collision cross-section. In this section, we introduce this reactive Boltzmann collision operator in the context of the reactive--elastic system discussed in \cite{Rossani_Spiga_1999}, and then specialize it to effectively irreversible exoergic reactions such as nuclear fusion.

Consider a spatially homogeneous system consisting of four different species undergoing the reversible reaction:
\begin{equation} \label{reaction}
A^1+A^2\leftrightarrow A^3+A^4.
\end{equation}
Let $f^{(i)}(t,\bv)$ ($i=1,\dots,4$) denote the distribution function of species $i$, depending on time $t\geq 0$ and velocity $\bv \in \mathbb{R}^3$. Each species is characterized by its particle mass $m_i$ and internal energy $U_i$.  Without loss of generality, we assume that $U_1+U_2\leq U_3+U_4$. The system is governed by
\begin{equation} \label{Boltzmann}
\partial_t f^{(i)}=Q_i^R[\underline{f}]+\sum_{j=1}^4 Q^E[f^{(i)},f^{(j)}], \quad i=1,\dots, 4,
\end{equation}
where $\underline{f}:=\{f^{(1)},f^{(2)},f^{(3)},f^{(4)}\}$, $Q_i^R$ is the reactive collision operator acting on species $i$, and $Q^E[f^{(i)},f^{(j)}]$ is the elastic collision operator between species $i$ and $j$.

The reactive collision operator $Q_i^R$ is defined by 
\begin{align} \label{react}
Q_{i}^R[\underline{f}](\bv)&=\int_{\mathbb{R}^3} \int_{S^2} \Theta(|\bv-\bw|^2-\delta_{ij}^{hk}) \left[B_{hk}^{ij}(|\bv_{ij}^{hk}-\bw_{ij}^{hk}|,\cos\theta)\left(\frac{\mu_{ij}|\bv_{ij}^{hk}-\bw_{ij}^{hk}|}{\mu_{hk}|\bv-\bw|} \right)f^{(h)}(\bv_{ij}^{hk})f^{(k)}(\bw_{ij}^{hk})\nonumber\right.\\
& \hspace{2.5in} \left.-B_{ij}^{hk}(|\bv-\bw|,\cos\theta) f^{(i)}(\bv)f^{(j)}(\bw)\right]\rd{\bm \sigma}\rd{\bw}\nonumber\\
:&=Q_i^{R,+}[f^{(h)},f^{(k)}]-Q_i^{R,-}[f^{(i)},f^{(j)}],
\end{align}
where the species indices $(i,j,h,k)$ take one of the tuples $(1,2,3,4)$, $(2,1,4,3)$, $(3,4,1,2)$, or $(4,3,2,1)$. In \eqref{react}, the pre- and post-collision velocities satisfy conservation of momentum and internal plus kinetic energy:
\begin{equation}\label{conservation}
\begin{aligned}
m_h\bv_{ij}^{hk}+m_k\bw_{ij}^{hk}&=m_i\bv+m_j\bw,\\
m_h|\bv_{ij}^{hk}|^2+2U_h+m_k|\bw_{ij}^{hk}|^2+2U_k&=m_i|\bv|^2+2U_i+m_j|\bw|^2+2U_j.
\end{aligned}
\end{equation}
Let $\bm \sigma$ denote the direction of $\bv_{ij}^{hk}-\bw_{ij}^{hk}$. The collision kernel $B_{ij}^{hk}$ is given by
\begin{equation}\label{eq:B_ijhk}
B_{ij}^{hk}(|\bv-\bw|,\cos \theta) = |\bv-\bw|\sigma_{ij}^{hk}(|\bv-\bw|,\cos\theta), \quad \cos \theta=\bm \sigma\cdot \frac{\bv-\bw}{|\bv-\bw|},
\end{equation}
with $\sigma_{ij}^{ij}$ being the differential cross-section and $\theta$ the scattering angle. Finally, the function $\Theta$ in \eqref{react} is the Heaviside step function enforcing the reaction energy threshold, and
\begin{equation}
\delta_{ij}^{hk}:=\frac{2(U_h+U_k-U_i-U_j)}{\mu_{ij}},
\end{equation}
where $\mu_{ij}=m_im_j/(m_i+m_j)$ is the reduced mass. The principle of microscopic reversibility \cite{Light_Ross_Shuler_1969}, based on the time symmetry of the Schr\"{o}dinger equation or the classical Liouville equation, can be used to rewrite the forward reaction collision kernel in terms of the backward reaction and vice versa. Equation \eqref{react} can then be written as 
\begin{equation}
\begin{aligned}
Q_i^R[\underline{f}](\bv)
={}&
\int_{\mathbb{R}^3}\int_{S^2}
\Theta\!\left(|\bv-\bw|^2-\delta_{ij}^{hk}\right)
B_{ij}^{hk}\!\left(|\bv-\bw|,\cos\theta\right)
\Bigg[
\left(\frac{\mu_{ij}}{\mu_{hk}}\right)^3
f^{(h)}(\bv_{ij}^{hk})f^{(k)}(\bw_{ij}^{hk})
\\
&\hspace{3.0in}
-f^{(i)}(\bv)f^{(j)}(\bw)
\Bigg]
\,\rd{\bm{\sigma}}\,\rd{\bw}.
\end{aligned}
\end{equation}
Although these formulations are equivalent, we use the formulation in \eqref{react} throughout the remainder of the paper because, for fusion reactions, it is expressed in terms of the forward-reaction collision cross-section, for which established parameterizations exist.

Note that if $(i,j) = (h,k)$, then $\mu_{hk} = \mu_{ij}$ and $\delta_{ij}^{hk} = 0$. Thus, one recovers the classical Boltzmann collision operator \cite{Cercignani} for elastic collisions: 
\begin{equation} \label{elastic}
Q^E[f^{(i)},f^{(j)}](\bv)= \int_{\mathbb{R}^3} \int_{S^2}B_{ij}^{ij}(|\bv-\bw|,\cos\theta)\left[f^{(i)}(\bv_{ij}^{ij})f^{(j)}(\bw_{ij}^{ij})-f^{(i)}(\bv)f^{(j)}(\bw)\right]\rd{\bm \sigma}\rd{\bw},
\end{equation}
where the pre- and post-collision velocities again satisfy conservation of momentum and energy:
\begin{equation}
m_i\bv_{ij}^{ij}+m_j\bw_{ij}^{ij}=m_i\bv+m_j \bw, \quad m_i|\bv_{ij}^{ij}|^2+m_j|\bw_{ij}^{ij}|^2=m_i|\bv|^2+m_j |\bw|^2,
\end{equation}
and the collision kernel $B_{ij}^{ij}$ is defined analogously to \eqref{eq:B_ijhk}.

\subsection{Specialization to irreversible exoergic reactions}

The fusion reactions considered in this work are treated as effectively irreversible, since the inverse reaction is neglected in the kinetic model. Under this approximation, consider the exoergic reaction
\begin{equation} \label{irreaction}
A^3+A^4 \to A^1+A^2.
\end{equation}
Then, the system \eqref{Boltzmann} reduces to the following form: 
\begin{equation}\label{irreact}
\begin{aligned}
&\partial_t f^{(1)}= Q_1^{R,+}[f^{(3)}, f^{(4)}] +\sum_{j=1}^4 Q^E[f^{(1)},f^{(j)}],  \quad
\partial_t f^{(2)}=  Q_2^{R,+}[f^{(4)}, f^{(3)}] +\sum_{j=1}^4 Q^E[f^{(2)},f^{(j)}],  \\
&\partial_t f^{(3)}= -Q_3^{R,-}[f^{(3)}, f^{(4)}] +\sum_{j=1}^4 Q^E[f^{(3)},f^{(j)}],  \quad
\partial_t f^{(4)}= -Q_4^{R,-}[f^{(4)}, f^{(3)}] +\sum_{j=1}^4 Q^E[f^{(4)},f^{(j)}], \end{aligned}
\end{equation}
where $Q_1^{R,+}[f^{(3)}, f^{(4)}]$ and $Q_2^{R,+}[f^{(4)}, f^{(3)}]$ are given by the gain term of \eqref{react}, and $Q_3^{R,-}[f^{(3)}, f^{(4)}]$ and $Q_4^{R,-}[f^{(4)}, f^{(3)}]$ are given by the loss term of \eqref{react}. Note that, due to the exoergic nature of the reaction, $\delta_{34}^{12}$ and $\delta_{43}^{21}$ are both non-positive. Therefore, the Heaviside function is identically equal to 1 in $Q_3^{R,-}[f^{(3)}, f^{(4)}]$ and $Q_4^{R,-}[f^{(4)}, f^{(3)}]$ and can be omitted from the collision integral. 

For the neutron branch of deuterium--deuterium fusion considered in Sec.~\ref{ss:numerical_eg}, we set $f^{(1)} = f^{(He)}$, $f^{(2)} = f^{(n)}$, $f^{(3)} = f^{(4)} = f^{(D)}$. If we neglect the species $f^{(2)}$ and take into account that the species $f^{(3)}$ and $f^{(4)}$ are artificial copies of the same species, the above system reduces to
\begin{equation}
\begin{aligned} \label{irreact1}
&\partial_t f^{(He)}= \frac{1}{2}Q_{He}^{R,+}[f^{(D)}, f^{(D)}] + Q^E[f^{(He)},f^{(D)}]+Q^E[f^{(He)},f^{(He)}],\\
&\partial_t f^{(D)}= -Q_D^{R,-}[f^{(D)}, f^{(D)}] + Q^E[f^{(D)},f^{(D)}]+Q^E[f^{(D)},f^{(He)}].
\end{aligned}
\end{equation}
The fusion reactivity is obtained by integrating the deuterium reactive loss term over velocity and normalizing by the product of the reactant number densities:
\begin{equation}
\label{fureact}
\langle \sigma v \rangle
:=\frac{1}{n_D^2}
\int_{\mathbb{R}^3}
Q_D^{R,-}[f^{(D)},f^{(D)}](\bm v)\rd \bm v=
\frac{1}{n_D^2}
\int_{\mathbb{R}^3}\!\int_{\mathbb{R}^3} 
|\bv-\bw| \, \Sigma_{DD}^{He \, n}(|\bv-\bw|)\,
f^{(D)}(\bm v)\,f^{(D)}(\bm w)
\rd \bm v \rd \bm w,
\end{equation}
where $\Sigma_{DD}^{He \, n}$ is the total cross-section given by
\begin{equation}
\Sigma_{DD}^{He \, n}(|\bv-\bw|)
=
\int_{S^2}
\sigma_{DD}^{He n}\left(|\bv-\bw|, \sigma\cdot \frac{\bv-\bw}{|\bv-\bw|}\right)
\rd \sigma=2\pi \int_0^{\pi} \sigma_{DD}^{He n}(|\bv-\bw|,\cos \theta) \sin \theta \rd{\theta},
\end{equation}
and $n_D$ is the number density
\begin{equation}
n_D = \int_{\mathbb{R}^3} f^{(D)}(\bm v)\rd \bv.
\end{equation}

\section{Variants of the multi-species elastic collision operator}

In the presence of suprathermal heating, fusion fuel velocity distribution functions have been speculated to attain a non-Maxwellian quasi-steady state \cite{Ye_Zhang_Wan_PoP_2025, Xie2023_fusion_reactivities}, in which high-velocity tail depletion from fusion reactions is balanced by rapid thermalization due to Coulomb collisions. Capturing these dynamics self-consistently thus requires a high-fidelity treatment of both reactive and elastic interactions. In this section, we discuss the elastic collision operators considered in this work.

We consider three types of elastic collision operators: the classical Boltzmann operator \eqref{elastic} with a cut-off Coulomb collision kernel, the Landau operator, and the LB operator. The Landau operator can be viewed as an approximation of the Boltzmann operator in the grazing-collision limit, while the LB operator is a further simplified approximation of the Landau operator.

\subsection{The Boltzmann operator with cut-off Coulomb collision kernel and its Landau asymptotics}\label{ss:theory_boltzC}

The unscreened Coulomb collision kernel in \eqref{elastic} has a non-integrable singularity in the scattering angle $\theta$:
\begin{equation}\label{eq:unscreened_coulomb_kernel}
B_{ij}^{ij}(|\bm \bv-\bw|, \cos \theta)
=
\frac{Z_i^2 Z_j^2 e^4 (m_i + m_j)^2}{64 \pi^2 \epsilon_0^2 (m_i m_j)^2}
\cdot
\frac{1}{|\bv-\bw|^3 \sin^4(\theta/2)},
\end{equation}
where $Z_i$ is the charge number of the species $i$, $e$ is the fundamental charge, and $\epsilon_0$ is the vacuum permittivity. To make the Boltzmann--Coulomb collision operator well-posed, one must truncate contributions from interactions with scattering angle smaller than a cutoff $\varepsilon$. In plasma physics, $\varepsilon$ is typically chosen so that collisions with impact parameter larger than the Debye length $\lambda_D$ are excluded. Expressing the scattering angle in terms of the impact parameter and approximating the relative kinetic energy by a characteristic thermal energy, $ \mu_{ij}|\bv-\bw|^2 \sim \frac{3}{2}(T_i + T_j)$ (where $T_i$ is the temperature of species $i$ in energy units) yields \cite{Degond_monograph_2007}:
\begin{equation}\label{eq:coulomb_cutoff}
\varepsilon
=
\frac{Z_i Z_j e^2}{3\pi \epsilon_0 (T_i + T_j)\lambda_D}.
\end{equation}
Note that $\varepsilon$ is the inverse of the plasma parameter, $\Lambda$, such that $|\ln \varepsilon| =  |\ln \Lambda|$ is the Coulomb logarithm. With this, the cut-off elastic collision operator \eqref{elastic} becomes
\begin{equation}\label{cutoffBoltz}
Q_\varepsilon^B[f^{(i)},f^{(j)}](\bm v)
=
\int_{\mathbb{R}^3}
\int_{0}^{2\pi}
\int_{\varepsilon}^{\pi}
B_{ij}^{ij}(|\bv-\bw |,\cos\theta)
\left[
f^{(i)}(\bm v_{ij}^{ij}) f^{(j)}(\bm w_{ij}^{ij})
-
f^{(i)}(\bm v) f^{(j)}(\bm w)
\right]\sin \theta
\rd \theta \rd \varphi \rd \bm w.
\end{equation}
It has been shown in \cite{Degond_Desreux_FP_asymptotics} that the leading order term in the $\varepsilon \to 0$ asymptotics of \eqref{cutoffBoltz} yields the multi-species Landau operator:
\begin{equation}
\label{Landau}
Q^L[f^{(i)},f^{(j)}](\bm v)
= \nabla_{\bm v} \cdot 
\int_{\mathbb{R}^3}
A_{ij}(\bv-\bw)
\left(
\frac{1}{m_i}
f^{(j)}(\bm w)\,\nabla_{\bm v} f^{(i)}(\bm v)
- \frac{1}{m_j} f^{(i)}(\bm v)\,\nabla_{\bm w} f^{(j)}(\bm w)
\right)
\rd \bm w,
\end{equation}
where the collision kernel, $A_{ij}$, projects onto directions orthogonal to the relative velocity:
\begin{equation}\label{landau_kernel}
    A_{ij}(\bm z) = C_{\gamma} \, \frac{|\bm z|^{\gamma + 2}}{m_i} \left(
\bm{I} - \frac{\bm z \, \otimes \, \bm z}{|\bm z|^2}
\right).
\end{equation}
In the case of Coulomb scattering, $\gamma = -3$, the leading constant is given by
\begin{equation}\label{eq:landau_phys_constant}
C_{-3} = \frac{Z_i^2 Z_j^2 e^4 \, |\ln\varepsilon|}{8 \pi \epsilon_0^2}.
\end{equation}

\subsection{The multi-species Lenard--Bernstein operator}\label{ss:MLB_theory}

We adopt the multi-species LB model proposed in \cite{HHHH_2025}, which conserves mass, momentum, and energy, and satisfies the $\mathcal{H}$-theorem, as does the full multi-species Landau operator \eqref{Landau}. In addition, the model parameters can be chosen to simultaneously match the momentum and temperature relaxation rates derived from the cut-off Boltzmann--Coulomb operator for a binary mixture in a near-equilibrium regime.

The model is given by
\begin{equation}\label{MLB}
Q^{LB}[f^{(i)},f^{(j)}](\bm v)
=
\lambda_{ij}\,\frac{T_{ij}}{m_i}\,
\nabla_{\bm v}\cdot
\left[
M^{(ij)}(\bm v)\,
\nabla_{\bm v}
\left(
\frac{f^{(i)}(\bm v)}{M^{(ij)}(\bm v)}
\right)
\right],
\end{equation}
where \(\lambda_{ij}\) is a velocity-independent collision frequency between species $i$ and $j$, and $M^{(ij)}$ is the mixture Maxwellian defined by
\begin{equation}
M^{(ij)}(\bv)
:=
n_i\left(\frac{m_i}{2\pi T_{ij}}\right)^{3/2}
\exp\left(-\frac{m_i|\bv-\bu_{ij}|^2}{2T_{ij}}\right).
\end{equation}
Here, the species mass density, bulk velocity, temperature, and energy are defined by
\begin{equation}
\begin{aligned}
&\rho_i = m_i n_i = m_i \int_{\mathbb{R}^3} f^{(i)}(\bm v)\rd \bm v,
\quad
\bm u_i = \frac{1}{n_i} \int_{\mathbb{R}^3} \bm v\, f^{(i)}(\bm v)\rd \bm v, \\
&T_i = \frac{m_i}{3 n_i} \int_{\mathbb{R}^3} |\bm v - \bm u_i|^2 f^{(i)}(\bm v)\rd \bm v, \quad
E_i = \frac{1}{2} \int_{\mathbb{R}^3} m_i |\bm v|^2 f^{(i)}(\bm v) \rd \bm v=\frac{1}{2}\rho_i |\bu_i|^2+\frac{3}{2}n_iT_i,
\end{aligned}
\end{equation}
and the pairwise velocity and temperature are defined by 
\begin{equation}
\bm u_{ij}
=
\alpha_{ij} \bm u_i+(1-\alpha_{ij}) \bm u_j,
\quad
T_{ij}
=
\beta_{ij} T_i+(1-\beta_{ij}) T_j
+
\frac{\gamma_{ij}}{3}\,|\bm u_i - \bm u_j|^2,
\end{equation}
where the weights are free parameters chosen to match the properties of the Boltzmann collision operator. We take 
\begin{equation}
\alpha_{ij}
=
1 - \frac{1}{2} \frac{m_i + m_j}{m_i} \frac{\xi_{ij}}{n_i \lambda_{ij}},
\quad
\beta_{ij}
=1-\frac{m_i}{m_i+m_j}(1-\alpha_{ij}),
\quad
\gamma_{ij}
=\frac{m_i m_j}{m_i+m_j}(1-\alpha_{ij}),
\end{equation}
where\footnote{It is common in the literature (\cite{HHHH_2025, Morse_PhysFluids_1963}) to approximate $| \ln(\sin(\varepsilon/2))|$ by $|\ln \varepsilon|$ in the definition of $\xi_{ij}$, since the difference is small for $\varepsilon \ll 1$.}
\begin{equation}
    \xi_{ij}
    =
    \frac{2}{3 (2 \pi)^{3/2}}
    \frac{
    Z_i^2 Z_j^2 e^4
        n_i n_j \left| \ln(\sin(\varepsilon/2)) \right|
    }{
        \epsilon_0^2 m_i m_j
        \left( \frac{T_i}{m_i} + \frac{T_j}{m_j} \right)^{3/2}
    },
    \quad
    \lambda_{ij}
    =
    \frac{\xi_{ij}}{n_i} \,
    \max_{i,j} \left( 
    \frac{m_i + m_j}{m_i}
    \right),
\end{equation}
with the maximum taken over all interacting species. Notice that cross-species interactions in \eqref{MLB} are entirely mediated through the mixture Maxwellian. Thus, the details of highly non-Maxwellian cross-species interactions are necessarily missed by this model.

\section{Numerical methods for time-dependent reactive collisional kinetics}

In this section, we describe the numerical methods used to discretize the irreversible reactive--elastic system \eqref{irreact}, in which the reactive operator is given by either the gain or the loss term of \eqref{react}, and the elastic operator is taken to be the cut-off Boltzmann--Coulomb operator \eqref{cutoffBoltz}, the Landau operator \eqref{Landau}, or the LB operator \eqref{MLB}.

\subsection{The fast Fourier spectral method for the reactive Boltzmann operator}\label{ss:BoltzR_spectral_method}


First, we introduce notation for the pre-- and post--collision relative velocities, $\bm q := \bm v - \bm w$ and $\bm q_{ij}^{hk} := \bm v_{ij}^{hk} - \bm w_{ij}^{hk}$. Then the angular integration variable in \eqref{react}, $\bm \sigma$, equals $\hat{\bm q}_{ij}^{hk}$, the direction of $\bm q_{ij}^{hk}$. Following the conservation laws in \eqref{conservation}, we have
\begin{equation}
\bm v_{ij}^{hk}
=
\bm v-\frac{m_j}{M}\bm q+\frac{m_k}{M} |\bm q_{ij}^{hk}| \hat{\bm q}_{ij}^{hk},\quad
\bm w_{ij}^{hk}
=
\bm v-\frac{m_j}{M}\bm q-\frac{m_h}{M} |\bm q_{ij}^{hk}| \hat{\bm q}_{ij}^{hk},
\end{equation}
where $M=m_i+m_j=m_h+m_k$, and, by conservation of internal plus kinetic energy,
\begin{equation}
\label{qq-relation}
|\bm q_{ij}^{hk}|
=
\left[
\frac{\mu_{ij}}{\mu_{hk}}
\left(|\bm q|^2-\delta_{ij}^{hk}\right)
\right]^{1/2}.
\end{equation}
With this notation and a change of integration variable $\bw\rightarrow \bm q$, the gain and loss terms of the collision operator \eqref{react} become, respectively,
\begin{align}
\label{truncatedQgain}
Q_i^{R,+}[f^{(h)},f^{(k)}](\bv)&=\int_0^{R^+} \int_{S^2}\int_{S^2} \Theta(|\bm q|^2-\delta_{ij}^{hk}) B_{hk}^{ij}(|\bm q_{ij}^{hk}|, \bm{\hat q}_{ij}^{hk}  \cdot \bm{\hat q})\left(\frac{\mu_{ij}|\bm q_{ij}^{hk}|}{\mu_{hk}|\bm q|} \right) \nonumber\\
&\hspace{2.25in} \times f^{(h)}(\bv_{ij}^{hk})f^{(k)}(\bw_{ij}^{hk})|\bm q|^2\rd{\hat{\bm q}_{ij}^{hk}}\rd{\hat{\bm q}}\rd{|\bm q|},
\end{align}
\begin{equation}
\label{truncatedQloss}
Q_i^{R,-}[f^{(i)},f^{(j)}](\bv)=\int_0^{R^-} \int_{S^2}  \int_{S^2} \Theta(|\bm q|^2-\delta_{ij}^{hk}) B_{ij}^{hk}(|\bm q|,\bm{\hat q}_{ij}^{hk}  \cdot \bm{\hat q}) f^{(i)}(\bv)f^{(j)}(\bv -\bm q)|\bm q|^2\rd{\hat{\bm q}_{ij}^{hk}}\rd{\hat{\bm q}}\rd{|\bm q|}.
\end{equation}
Note that we have also truncated the integrals of $\bm q$ above to the balls of radii $R^{\pm}$ centered at the origin, $\mathcal{B}(0,R^{\pm})$. Assuming that the supports of the reactant velocity distribution functions are contained in $\mathcal{B}(0,S^-)$, it suffices to take
\begin{equation}\label{eq:domain_sizing_energy_cons}
R^+=\left[
4(S^-)^2\frac{\mu_{hk}}{\mu_{ij}}
+\delta_{ij}^{hk}
\right]^{1/2}, \quad R^-=2S^-.
\end{equation}

Next, we choose a computational domain $\mathcal{D}_L=[-L,L]^3$ and approximate the distribution function $f^{(i)}$ (resp., $f^{(j)}$, $f^{(h)}$, and $f^{(k)}$) by a truncated Fourier series in $\mathcal{D}_L$:
\begin{equation}
\label{truncatedFourier}
f^{(i)}(\bm v) \approx \sum\limits_{\bm k = -N/2}^{N/2 - 1}
\hat{f}^{(i)}_{\bm k}\,e^{\im \frac{\pi}{L} \bm k \cdot \bm v}.
\end{equation}
The domain size must be chosen large enough to control aliasing errors. For the numerical experiments in Sec.~\ref{ss:numerical_eg}, we follow the Landau operator domain-sizing convention \cite{PRT00} and take $L=2S$, assuming that the supports of all species involved are contained in $\mathcal{B}(0,S)$. This choice is sufficient for the elastic operators used in Sec.~\ref{ss:numerical_eg}, the Landau and LB operators. In our computations, this sizing was also adequate for the irreversible reactive Boltzmann terms, since they are determined by the reactant distributions, whose velocity support is much smaller than that of the fusion products involved in elastic interactions. For the Boltzmann operator with the cut-off Coulomb cross-section, used in Sec.~\ref{ss:elastic_relaxation_rate_matching}, we use the slightly more conservative sizing described in \cite{Jaiswal_Alexeenko_Hu}.

Plugging \eqref{truncatedFourier} into \eqref{truncatedQgain} and performing a Galerkin projection, the $\bm k$-th Fourier mode of the gain term can be written as
\begin{equation}
\label{fourier-gain}
\hat{Q}_{i, {\bm k}}^{R,+}
=
\sum_{\substack{\bm l, \bm m=-N/2 \\ \bm l + \bm m = \bm k}}^{N/2-1}
G_{hk}^{ij,+}(\bm l, \bm m)\,\hat{f}^{(h)}_{\bm l}\,\hat{f}^{(k)}_{\bm m},
\end{equation}
where the weight is
\begin{equation}
\label{weightgain}
G_{hk}^{ij,+}(\bm l, \bm m)
=
\int_0^{R^+}
\int_{S^2}
F_{hk}^{ij}(\bm l + \bm m,|\bm q|, \bm{\hat q}_{ij}^{hk})\Theta(|\bm q|^2-\delta_{ij}^{hk})
\left(\frac{\mu_{ij} |\bm q_{ij}^{hk}|}{\mu_{hk} |\bm q|}\right)
e^{\im\frac{\pi}{L}|\bm q_{ij}^{hk}|
\left(
\frac{m_k}{M} \bm l - \frac{m_h}{M} \bm m
\right)\cdot \bm{\hat q}_{ij}^{hk}}
|\bm q|^2
\,\mathrm{d} \bm{\hat q}_{ij}^{hk}\,\mathrm{d}|\bm q|,
\end{equation}
with
\begin{equation}
F_{hk}^{ij}(\bm l + \bm m,|\bm q|, \bm{\hat q}_{ij}^{hk})
=
\int_{S^2}
B_{hk}^{ij}(|\bm q_{ij}^{hk}|, \bm{\hat q}_{ij}^{hk}\cdot \hat{\bm q})
\,e^{-\im\frac{\pi}{L}|\bm q|\frac{m_j}{M}(\bm l + \bm m)\cdot \bm{\hat q}}
\,\mathrm{d} \bm{\hat q},
\end{equation}
where $|\bm q_{ij}^{hk}|$ is expressed in terms of $|\bm q|$ using \eqref{qq-relation}.

The loss term is given by
\begin{equation} \label{fourier-loss}
\hat{Q}_{i, {\bm k}}^{R,-}
=
\sum_{\substack{\bm l, \bm m=-N/2 \\ \bm l + \bm m = \bm k}}^{N/2-1}
G_{ij}^{hk,-}( \bm m)\,\hat{f}^{(i)}_{\bm l}\,\hat{f}^{(j)}_{\bm m},
\end{equation}
where the weight is
\begin{equation}\label{weightloss}
    G_{ij}^{hk,-}(\bm m) = \int_0^{R-}\int_{S^2}\int_{S^2} \Theta(|\bm q|^2-\delta_{ij}^{hk}) B_{ij}^{hk}(| \bm q|, \bm{\hat{q}}_{ij}^{hk}\cdot  \hat{\bm{q}}) \, e^{-\im \frac{\pi}{L} |\bm q|\bm m \cdot \hat{\bm q}} |\bm q|^2\rd \bm{\hat{q}}_{ij}^{hk} \rd \hat{\bm q}\rd |\bm q|.
\end{equation}

The loss term \eqref{fourier-loss} is readily written as a convolution, and can therefore be accelerated using the fast Fourier transform (FFT) with a computational complexity of $\mathcal{O}(N^3\log N)$. However, the gain term \eqref{fourier-gain} is a weighted convolution. Without further treatment, it must be evaluated directly, with a complexity $\mathcal{O}(N^6)$. To accelerate the computation of the gain term, we seek a low-rank approximation of the form
\begin{equation}\label{eq:low_rank_G3412_gain}
G_{hk}^{ij,+}(\bm l, \bm m)\approx \sum_{p=1}^{N_p}\alpha_p(\bm l + \bm m)\,\beta_p(\bm l)\,\gamma_p(\bm m),
\end{equation}
which turns the weighted convolution \eqref{fourier-gain} into a sum of $N_p$ pure convolutions, each of which can be evaluated with the FFT. This reduces the computational complexity from $\mathcal{O}(N^6)$ to $\mathcal{O}(N_p  N^3 \log N)$. As in \cite{GHHH_2017, Jaiswal_Alexeenko_Hu}, we construct this low-rank approximation from a numerical quadrature for \eqref{weightgain}, taking an $N_{|\bm q|} = \mathcal{O}(N)$ point Gauss-Legendre rule in the radial direction and an $N_{\bm{\hat q}_{ij}^{hk}} \ll \mathcal{O}(N^2)$ point spherical design over the unit sphere, such that $N_p = N_{|\bm q|} \cdot N_{\bm{\hat q}_{ij}^{hk}} \ll N^3$. 

The weights $F_{hk}^{ij}$ and $G_{ij}^{hk,-}$ are precomputed to high accuracy. For angularly-independent collision kernels, $F_{hk}^{ij}$ has an analytic form:
\begin{equation}\label{eq:F_analytic}
    F_{hk}^{ij}(\bm k, |\bm q|) = 4 \pi  B_{hk}^{ij}(|\bm q_{ij}^{hk}|) \ \mathrm{Sinc}
    \left(
    \frac{\pi |\bm k| |\bm q|}{L}  \frac{m_j}{M}
    \right).
\end{equation}
For fusion reactions, we employ the angularly-independent form above together with the Bosch--Hale parametrization~\cite{Bosch_Hale_1992} of the differential cross-section,
\begin{equation}\label{eq:bosch_hale}
B_{hk}^{ij}(|\bm q_{ij}^{hk}|)
=
|\bm q_{ij}^{hk}|\,
\frac{S(E_{ij}^{hk})}{E_{ij}^{hk}}
\exp\!\left(-\frac{B_G}{\sqrt{E_{ij}^{hk}}}\right),
\end{equation}
where $E_{ij}^{hk} = \tfrac{1}{2}\mu_{hk}|\bm q_{ij}^{hk}|^2$ is the center-of-mass energy of the reactants and $B_G$ is the Gamow constant. The exponential factor captures tunneling through the Coulomb barrier, while the $S$-factor encapsulates the nuclear interaction and varies slowly with energy. The function $S(E_{ij}^{hk})$ is approximated by Pad\'{e} polynomials, with coefficients provided in~\cite{Bosch_Hale_1992}. 

Specializing to irreversible exoergic systems of the form \eqref{irreact}, the kernels in the gain \eqref{eq:F_analytic} and loss \eqref{weightloss} weights correspond to the same reactive channel. Thus, $B_{ij}^{hk}(| \bm q|)$ in \eqref{weightloss} is simply a permutation of the indices in \eqref{eq:bosch_hale}.


\subsection{The fast Fourier spectral method for the cutoff Boltzmann--Coulomb operator} \label{ss:BoltzC_spectral_method}

The numerical method in the previous section does not generalize to Coulomb collisions because the weights \eqref{weightgain} and \eqref{weightloss} become divergent integrals, owing to the singularity in the unscreened Coulomb collision kernel \eqref{eq:unscreened_coulomb_kernel} in both the scattering angle and relative velocity. Instead, we use a fast Fourier spectral method in the spirit of \cite{Hu_Qi_noncutoff_BE}, generalized here to the multi-species elastic Boltzmann collision operator with the cutoff Coulomb cross-section, \eqref{cutoffBoltz}. This form treats the gain and loss term of the collision operator together, which provides some cancellation of the leading-order singular terms. Due to its high computational cost, we restrict its use to the verification of the Landau and LB operators in Sec.~\ref {ss:verification}, rather than full reactive--elastic simulations. 

We truncate the collision integral, represent the distribution functions with a truncated Fourier series, and project onto the Fourier basis, as before. After some algebra, we obtain the $\bm k$-th Fourier mode of the classical Boltzmann collision operator:
\begin{equation}\label{eq:kth_mode_BoltzC}
    \hat{Q}_{i, {\bm k}}^{B}
=
\sum_{\substack{\bm l, \bm m=-N/2 \\ \bm l + \bm m = \bm k}}^{N/2-1}
G_{ij}^{ij}( \bm l, \bm m)\,\hat{f}^{(i)}_{\bm l}\,\hat{f}^{(j)}_{\bm m}, 
\end{equation}
with the convolution weight:
\begin{equation}\label{eq:G_weak_multispecies}
    G_{ij}^{ij}(\bm l, \bm m)
=
\int_{0}^R
\int_{S^2}
\Phi_{ij}^{ij} \left(|\mathbf q|\right)\,
F_{ij}^{ij}(\bm l + \bm m, |\bm q|, \hat{\bm q}) \, 
e^{-\im\frac{\pi}{L}|\mathbf q|\, \bm m \cdot \hat{\mathbf q}}
| \bm q|^2
\rd \hat{\bm q}
\rd |\bm q|,
\end{equation}
where $R = 2S$, assuming that the species involved have velocities contained in $B(0,S)$. As before, $\bm \sigma$ and $\hat{\bm q}$ are unit vectors. 
\begin{equation}\label{eq:F_weak_multispecies}
\begin{aligned}
F_{ij}^{ij}(\bm k, |\bm q|, \hat{\bm q}) &=
\int_{S^2} b_{ij}^{ij}(\cos \theta) \left(
e^{\im\frac{ \pi}{L} \frac{m_j}{m_i + m_j} |\bm q| \bm k \cdot (\hat{\bm q} - \bm \sigma)} -1
\right) \rd \bm \sigma \\
&=
\int_0^{2 \pi}
\int_0^{\pi}
b_{ij}^{ij}(\cos \theta)
\left(
e^{\im\frac{ \pi}{L} \frac{m_j}{m_i + m_j} |\bm q| \, \left( \bm k \cdot\hat{\bm q}(1 - \cos \theta) - |\bm k_{\perp}|\sin \theta \sin \phi 
\right)}
-1
\right) \sin \theta \rd \theta \rd \phi,
\end{aligned}
\end{equation}
where
\begin{equation}
    | \bm k_{\perp}| = \sqrt{|\bm k|^2 - (\bm k \cdot \hat{\bm q})^2}.
\end{equation}
We have also used that the collision kernel's kinetic and angular dependence is separable in our applications of interest:
\begin{equation}
    B_{ij}^{ij}(|\bm q|, \cos \theta) = \Phi_{ij}^{ij}(| \bm q|) \, b_{ij}^{ij}(\cos \theta).
\end{equation}

For Coulomb interactions, $\Phi_{ij}^{ij} \sim |\bm q|^{-3}$ and $b_{ij}^{ij} \sim \sin^{-4}(\theta/2)$. A Taylor expansion of \eqref{eq:F_weak_multispecies} yields a leading factor of $| \bm q|$, which, together with the radial Jacobian, cancels the kinetic singularity of the Coulomb kernel in \eqref{eq:G_weak_multispecies}.
 
The angular singularity is more restrictive. The weight \eqref{eq:F_weak_multispecies} is integrable for mildly singular kernels, which satisfy
\begin{equation}
\sin\theta \cdot  b(\cos\theta) \sim K \theta^{-1-\nu},
\qquad \theta\to 0,
\qquad 0\leq \nu < 2 .
\end{equation}
However, the unscreened Coulomb kernel corresponds to the edge case $\nu = 2$, for which the angular integral \eqref{eq:F_weak_multispecies} diverges. Thus, we truncate the integration domain to $\theta \in [\varepsilon, \pi]$ in the precomputation of \eqref{eq:F_weak_multispecies}, as discussed in Sec.~\ref{ss:theory_boltzC}.

As in the reactive gain term \eqref{fourier-gain}, the operator \eqref{eq:kth_mode_BoltzC} can be computed in $\mathcal{O}(N_p N^3 \log N)$ operations by using an $N_p$-point quadrature to evaluate \eqref{eq:G_weak_multispecies}. We observe that the required $N_p$ to obtain a given numerical accuracy is higher for the Coulomb kernel than for mildly singular or angularly independent kernels.

\subsection{The fast Fourier spectral method for the multi-species Landau operator}
\label{sec:Landau_spectral_method}

The fast Fourier spectral method for the multi-species Landau operator \eqref{Landau} follows a similar idea as originally proposed in \cite{PRT00} for the single-species case. After a change of integration variable $\bw\rightarrow \bm q=\bw-\bv$ and a truncation of the integral, we obtain
\begin{equation}
Q^L[f^{(i)}, f^{(j)}](\bm v)
= \nabla_{\bm v} \cdot \int_{\mathcal{B}(0,R)} A_{ij}(\bm q) \left(
\frac{1}{m_i}  
f^{(j)}(\bm v + \bm q) \nabla_{\bm v} f^{(i)}(\bm v) - \frac{1}{m_j} f^{(i)}(\bm v) \nabla_{\bm q}f^{(j)}(\bm v + \bm q)
    \right)
     \rd \bm q,
\end{equation}
where we take $R=2S$ again.

We again represent the distribution function by a truncated Fourier series \eqref{truncatedFourier} and perform a Galerkin projection. The resulting $\bm k$-th Fourier mode of the Landau operator has the form
\begin{equation}
\hat{Q}^L_{\bm k}
=
-\left(
\frac{\pi}{L}
\right)^2
\bm k^T
\sum_{\substack{\bm l, \bm m=-N/2 \\ \bm l + \bm m = \bm k}}^{N/2 - 1}
\left[
\frac{1}{m_i}
F_{ij}(\bm l) \bm m - \frac{1}{m_j}
F_{ij}(\bm l) \bm l
\right] \,
\hat{f}^{(j)}_{\bm l}
\hat{f}^{(i)}_{\bm m},
\end{equation}
where $\bm k^T$ denotes the transpose of the column vector $\bm k := [k_1; k_2; k_3]$, and
\begin{equation}\label{precomp_F_landau}
    F_{ij}(\bm k) := \int_{\mathcal{B}(0,R)} A_{ij}(\bm q) e^{\im \frac{\pi}{L} \bm k \cdot \bm q} \rd \bm q.
\end{equation}
Like the loss term of the reactive Boltzmann operator \eqref{fourier-loss}, this operator has a pure convolution form and can therefore be evaluated in $\mathcal{O}(N^3 \log N)$ operations via the FFT, provided that the weight $F_{ij}$ is precomputed to high accuracy.

\subsection{The Lenard--Bernstein operator discretization and time integration}

The primary advantage of the LB collision operator is that it admits an implicit time-integration strategy \cite{HHHH_2025}, which enables an implicit--explicit (IMEX) scheme for fusion-elastic systems. We illustrate this on reversible systems of the form \eqref{Boltzmann}, assuming that $Q^E=Q^{LB}$ as defined in \eqref{MLB}. The irreversible reactive case requires only replacing $Q_i^R[\underline{f}]$ with the appropriate irreversible gain or loss term, leaving the remainder of the formulation unchanged.

We employ a first-order IMEX discretization:
\begin{equation}\label{IMEX_general}
    \frac{f^{(i),n+1} - f^{(i),n}}{\Delta t}
    = Q_i^{R}[\underline{f}^{n}]
    + \sum_{j = 1}^{4}
    Q^{LB}\!\left[f^{(i),n+1}, f^{(j),n+1}\right],
\end{equation}
where the reactive term is treated explicitly, while the relatively stiff LB operator is treated implicitly. Since \(Q^{LB}\) is nonlinear only through the Maxwellian \(M^{(ij)}\), we first predict the value of \(M^{(ij)}\) at time $t^{n+1}$ using moment update equations, which effectively linearizes the collision operator. Specifically, taking the moments $\int \cdot \ m_i(1,\bv,|\bv|^2/2)\rd{\bv}$ on both sides of \eqref{IMEX_general} yields
\begin{equation}
\begin{aligned}
&\frac{\rho^{n+1}_i - \rho^{n}_i}{\Delta t}
=
\int_{\mathbb{R}^3} m_i Q_i^{R}[\underline{f}^{n}]\rd{\bv},\\
&\frac{\rho^{n+1}_i\bu^{n+1}_i - \rho^{n}_i\bu^{n}_i}{\Delta t}
=
\int_{\mathbb{R}^3} m_i \bm v \ Q_i^{R}[\underline{f}^{n}]\rd{\bv}
+
\sum_{j=1}^4
\rho_i^{n+1} \lambda_{ij}^{n+1}(1-\alpha_{ij})(\bm u_j^{n+1} - \bm u_i^{n+1}),\\
&\frac{E^{n+1}_i- E^{n}_i}{\Delta t}
=
\int_{\mathbb{R}^3}\frac12 m_i |\bm v|^2\ Q_i^{R}[\underline{f}^{n}]\rd{\bv}
+
3 \, \sum_{j=1}^4 n_i^{n+1} \lambda_{ij}^{n+1}(1-\beta_{ij})(T_j^{n+1}-T_i^{n+1})
\\
&\hspace{1in}
+
\sum_{j=1}^4
n_i^{n+1} \lambda_{ij}^{n+1}(1-\beta_{ij})
\Big[
m_i\, \bm u_i^{n+1}\cdot(\bm u_j^{n+1} - \bm u_i^{n+1})
-
m_j \, \bm u_j^{n+1}\cdot(\bm u_i^{n+1} - \bm u_j^{n+1})
\Big].
\end{aligned}
\end{equation}
The density update is explicit. The momentum and energy updates form a coupled nonlinear system, which we solve using the Gauss--Seidel-type iteration in \cite{HHW_2025, HHHH_2025}. 

Once \(\rho_i^{n+1}\), \(\bu_i^{n+1}\), and \(T_i^{n+1}\) are obtained, the Maxwellian \(M^{(ij),n+1}\) is fully determined, and the LB operator becomes effectively linear. We then discretize it in velocity space using a
central finite-difference scheme with no-flux boundary conditions. More details can be found in \cite{HHHH_2025}. Since we work in three-dimensional velocity space, the resulting system is quite large and takes the form 
\begin{equation}
\left( I -\Delta t (G_x + G_y + G_z) \right) f^{(i),n+1}
= f^{(i),n}
+
\Delta t\, Q_i^{R}[\underline{f}^{n}],
\end{equation}
where $G_x$, $G_y$, and $G_z$ denote the discretizations of the MLB operator in the corresponding velocity dimensions. To avoid solving a global $N^3$-dimensional linear system, we employ a first-order locally one-dimensional (LOD) splitting, yielding the sequence of systems
\begin{equation}\label{eq:IMEX_LOD_MLB}
\begin{aligned}
(I - \Delta t G_x)\, f^{*} &= f^{(i),n} + \Delta t\,Q_i^{R}[\underline{f}^{n}], \\
(I -\Delta t G_y)\, f^{**} &=  f^{*}, \\
(I - \Delta t G_z)\,  f^{(i),n+1} &= f^{**}.
\end{aligned}
\end{equation}

\section{Verification of collision operator calculations}\label{ss:verification}

\subsection{Spectral convergence of the Landau operator}

We verify the fast spectral method for the Landau operator with an analytic BKW-type solution to the spatially homogeneous two-species Landau equation 
\begin{equation}\label{elastic_system}
\partial_t f^{(i)}
=
\sum_{j=1}^{2} Q^L\!\left[f^{(i)}, f^{(j)}\right], \qquad i = 1,2,
\end{equation}
as presented in \cite{Carrillo_Hu_vanFleet_particle_Landau}. The solution exists when the Landau collision kernel is of the form \eqref{landau_kernel} with $\gamma = 0$, corresponding to the so-called Maxwell collision kernel:
\begin{equation}\label{eq:Landau_BKW_kernel}
A_{ij}(\bm z)
=
B_{ij}\left(|\bm z|^2 \bm I - \bm z \otimes \bm z\right),
\end{equation}
with the additional requirement that $B_{ij}$ is chosen such that
\begin{equation}
\beta_i := \sum_{j=1}^2 \frac{B_{ij}}{m_i m_j} n_j, \qquad \beta_1 = \beta_2 = \beta,
\end{equation}
for some constant $\beta$. We achieve this for $\beta = 1/16$ with arbitrary mass ratios by choosing
\begin{equation}
B_{11}=\frac{(m_1 / m_2)^2}{32},\qquad
B_{12}=B_{21}=\frac{(m_1 / m_2)}{32},\qquad
B_{22}=\frac{1}{32},
\end{equation}
and $n_1 = n_2 = 1$, as in \cite{Carrillo_Hu_vanFleet_particle_Landau}. The solution is given by
\begin{equation}\label{BKW_multispecies}
f_{\text{ext}}^{(i)}(t, \bm v)
=
n_i
\left(\frac{m_i}{2\pi K(t)}\right)^{3/2}
\exp\!\left(-\frac{m_i |\bm v|^2}{2K(t)}\right)
\left(
1
-
3\,Q(t)
+
\frac{m_i}{K(t)}\,Q(t)\,|\bm v|^2
\right),
\end{equation}

\begin{equation}\label{BKW_Landau_dfdt}
\begin{aligned}
\partial_t f_{\text{ext}}^{(i)}(t,\bm v)
&=
n_i
\left(\frac{m_i}{2\pi K(t)}\right)^{3/2}
\exp\!\left(-\frac{m_i |\bm v|^2}{2K(t)}\right)
\left[
\left(
\frac{m_i |\bm v|^2}{2K(t)^2}\dot K(t)
-
\frac{3}{2K(t)}\dot K(t)
\right)
\left(
1 - 3Q(t) + \frac{m_i}{K(t)} Q(t)\,|\bm v|^2
\right)
\right. \\
&\quad\left.
- 3\,\dot Q(t)
+ \frac{m_i}{K(t)} \dot Q(t)\,|\bm v|^2
- \frac{m_i}{K(t)^2} Q(t)\,\dot K(t)\,|\bm v|^2
\right], \qquad i=1,2,
\end{aligned}
\end{equation}
where  
\begin{equation}
Q(t)=\frac{1-K(t)}{2K(t)}, \qquad K(t) := 1 - C^{(BKW)} \, e^{-4 \beta t}. 
\end{equation}
We choose $C^{(BKW)} = 2/5$, which guarantees positivity of the solution at $t = 0$. 

To assess convergence, we evaluate the right-hand side of \eqref{elastic_system} with the fast spectral method from Sec.~\ref{sec:Landau_spectral_method} at time $t = 0.0$ with domain size $[-L,L]^3 = [-8,8]^3$, and we compute the absolute errors with respect to \eqref{BKW_Landau_dfdt} in the $L^{\infty}$ and Frobenius norms, respectively:
\begin{equation}
\epsilon_{L^\infty}^{(i)} := \left\| \partial_t f^{(i)} - \partial_t f^{(i)}_{\mathrm{ext}} \right\|_{L^\infty},
\qquad
\epsilon_{F}^{(i)} := \left\| \partial_t f^{(i)} - \partial_t f^{(i)}_{\mathrm{ext}} \right\|_{F}.
\end{equation}

\begin{table}[H]
\centering
\begin{tabular}{ccccc}
\hline
N & $\epsilon_{L^{\infty}}^{(1)}$ & $\epsilon_{L^{\infty}}^{(2)}$ & $\epsilon_{F}^{(1)}$ & $\epsilon_{F}^{(2)}$ \\
\hline
16 & 4.8973e-03 & 2.2483e-03 & 1.9251e-02 & 2.1598e-02 \\
32 & 1.4194e-04 & 4.8728e-06 & 9.3404e-04 & 4.0535e-05 \\
64 & 5.3929e-07 & 3.0762e-07 & 3.7682e-07 & 2.7801e-07 \\
\hline
\end{tabular}
\caption{Absolute errors for the multi-species Landau collision operator with
$L=8$, $m_1=1.5$, and $m_2=1.0$.}
\end{table}

\begin{table}[H]
\centering
\begin{tabular}{ccccc}
\hline
N & $\epsilon_{L^{\infty}}^{(1)}$ & $\epsilon_{L^{\infty}}^{(2)}$ & $\epsilon_{F}^{(1)}$ & $\epsilon_{F}^{(2)}$ \\
\hline
16 & 2.5789e-02 & 5.9234e-03 & 9.7761e-02 & 5.8988e-02 \\
32 & 1.6346e-02 & 9.4958e-05 & 3.9274e-02 & 2.9941e-04 \\
64 & 1.9004e-04 & 2.1184e-06 & 6.9808e-04 & 6.2308e-06 \\
\hline
\end{tabular}
\caption{Absolute errors for the multi-species Landau collision operator with
$L=8$, $m_1=5.0$, and $m_2=1.0$.}
\end{table}

The results show the expected rapid convergence for small mass ratios. At larger mass ratios, convergence degrades because the species distribution functions have increasingly disparate velocity-space scales, which are not equally resolved under finite resolution.

While this does not directly verify the physically relevant Coulomb case ($\gamma = -3$), the dependence on $\gamma$ enters only through the precomputed kernel factor \eqref{precomp_F_landau}, which can be precomputed to arbitrarily high accuracy.

\subsection{Elastic relaxation rate matching}\label{ss:elastic_relaxation_rate_matching}

We verify the consistency of the moment relaxation rates produced by the Boltzmann--Coulomb, Landau, and LB collision operators with a representative numerical example. All simulations use $N = 32$ velocity grid points per dimension, domain size $L = 1.5 \times 10^7$ m/s, and temporal step-size $\Delta t = 5 \times 10^{-10}$ s. For the Boltzmann--Coulomb operator we take $32$ quadrature points on the sphere and 32 quadrature points in the radial direction. We take species 1 to be deuterium and species 2 to be helium-3, with their physical masses. The initial condition consists of an $80\,\mathrm{keV}$ Maxwellian for species 1 and a Rosenbluth-type distribution \cite{Gamba_Haack_grazing_collisions} for species 2:
\begin{equation}\label{eq:T_relaxation_IC_maxw}
f^{(1)}(0,\bm{v}) =
\frac{n_0}{(2\pi T/m)^{3/2}}
\exp\!\left(
-\frac{m|\bm{v}|^2}{2T}
\right),
\end{equation}
\begin{equation}\label{eq:T_relaxation_IC_rosenbluth}
f^{(2)}(0,\bm{v}) =
C \,
\exp\!\left(
-10
\left(
\frac{|\bm{v}| - 0.3L}{0.3L}
\right)^2
\right),
\end{equation}
where $C$ is chosen to normalize the velocity-space integral of \eqref{eq:T_relaxation_IC_rosenbluth} to $n_0 = 10^{26} \mathrm{m}^{-3}$. This initial condition corresponds to a weakly coupled regime, with Coulomb logarithm $|\ln(\Lambda)| \approx 15$.

\begin{figure}[H]
\centering
\includegraphics{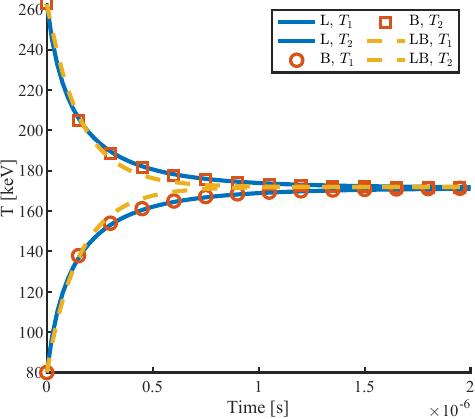}
\caption{
Temperature evolution of the Maxwellian--Rosenbluth initial condition subject to the spatially homogeneous Boltzmann equation with the Boltzmann--Coulomb (B) operator, Landau (L) operator, and Lenard--Bernstein (LB) operator. The Landau solution closely tracks the Boltzmann--Coulomb reference, while the Lenard--Bernstein model relaxes somewhat faster at early times when the distribution is furthest from Maxwellian equilibrium, as shown in Fig.~\ref{fig:T_relaxation_errors}.
}
\label{fig:T_relaxation}
\end{figure}

\begin{figure}[H]
    \centering
    \subcaptionbox{}[0.50\textwidth]
    {\includegraphics{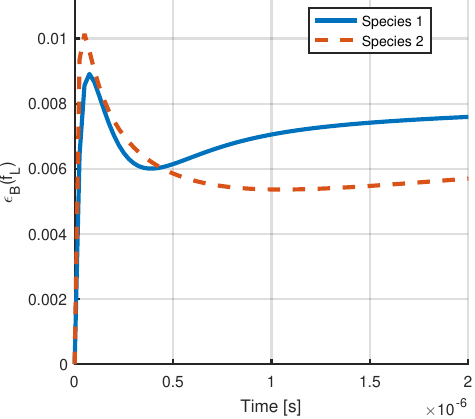}}
    \subcaptionbox{}[0.50\textwidth]
    {\includegraphics{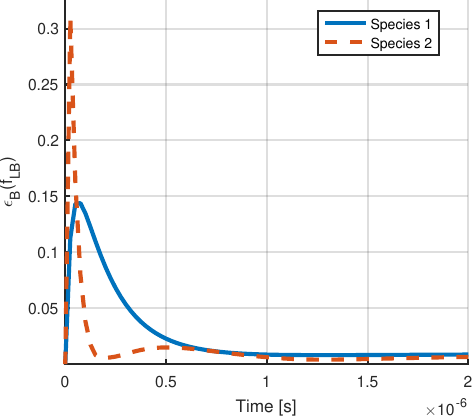}}
    \caption{
    Frobenius norm errors of the Landau (a) and LB (b) solutions relative to the Boltzmann--Coulomb solution. The relative error between the Boltzmann--Coulomb and Landau solutions is bounded by approx. 1\%, verifying the expected convergence in the given weakly coupled regime. The relative error between the Boltzmann and LB solutions is highest early in time, when the true solution deviates most from Maxwellian equilibrium, after which it decreases quickly as both models approach the correct steady state.
    }
    \label{fig:T_relaxation_errors}
\end{figure}

Figure~\ref{fig:T_relaxation} compares the temperature evolution predicted by the three collision models. By conservation of total energy, one can verify that the system approaches the analytic equilibrium temperature of $T_{\infty} = 171.6$ keV. In Fig.~\ref{fig:T_relaxation_errors}, we examine the similarity of the solutions to the different models with the relative Frobenius norm error with respect to the Boltzmann solution, $\epsilon_B(f) := ||f - f_B|| / ||f_B||$.

While the three operators produce qualitatively similar relaxation, the LB model relaxes noticeably faster than the Landau and Boltzmann-Coulomb operators at early to intermediate times, when the solution is furthest from Maxwellian equilibrium and the near-equilibrium Boltzmann-Coulomb relaxation rates used in the construction of \eqref{MLB} do not hold. This discrepancy has two distinct sources. First, the LB operator's collision frequency, $\lambda_{ij}$, is tuned to match the Boltzmann--Coulomb momentum and temperature relaxation rates (Sec.~\ref{ss:MLB_theory}) in a regime where both species' distribution functions remain close to a drifting Maxwellian. Second, the collision frequency, $\lambda_{ij}$ is velocity-independent, so even if it is tuned to match the moment relaxation rates in the regime of interest, it necessarily overestimates the thermalization rate of the high-velocity tail. 

\subsection{Grid convergence study of the reactive operator}\label{ss:grid_convergence}

We perform a grid refinement study using relative moment errors
\begin{equation}\label{eq:moment_error}
\epsilon_k(Q)
:=
\frac{
\left|m_k(Q)-m_k(Q_{\mathrm{ref}})\right|
}{
\left|m_k(Q_{\mathrm{ref}})\right|
},
\end{equation}
where $m_k( \cdot )$ denotes the $k$-th velocity moment:
\begin{equation}\label{eq:k_th_moment}
m_k(Q)
=
\int_{\mathbb{R}^3} Q(\bm v)\,|\bm v|^k \rd \bm v .
\end{equation}
Here $Q$ denotes either the irreversible deuterium loss operator $Q_{\mathrm D}^{R,-}$ or the irreversible helium-3 gain operator $Q_{\mathrm{He}}^{R,+}$, corresponding to the reactive loss and gain terms of the two-species system \eqref{irreact1}. The reference operators, $Q_{\mathrm{ref}}$, are both computed with $N=256$ Fourier modes per dimension. For the reference helium-3 gain operator, we use 192 quadrature points on the sphere and 256 points in the radial direction. The convergence studies use the fast spectral method described in Sec.~\ref{ss:BoltzR_spectral_method}, with velocity-domain size $L=2.8\times 10^7$ m/s, matching the setup of the numerical experiments in Sec.~\ref{ss:numerical_eg}.

\begin{figure}[H]
    \centering
    \subcaptionbox{\label{fig:loss_spectral}}{\includegraphics{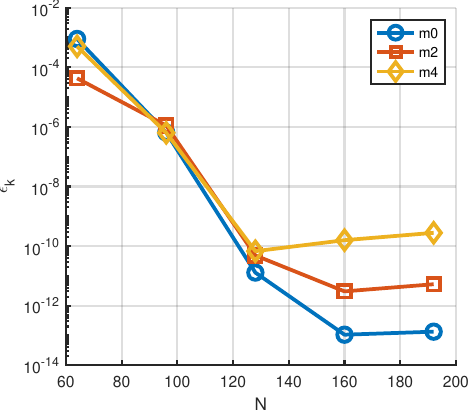}}%
    \subcaptionbox{\label{fig:gain_error}}{\includegraphics{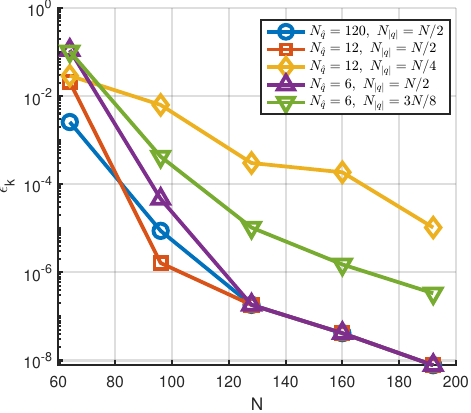}}
    \caption{(a) Spectral convergence of the loss operator measured using moment errors. (b) Spectral convergence of the zeroth moment of the gain operator for various quadrature configurations. Higher-order moments of the gain term behave similarly but are omitted for brevity. The choice of quadrature nodes $(N_{\hat{q}}, N_{|q|}) = (6,N/2)$ does not pollute the rapid convergence of the gain term past $N = 128$ and is therefore used in this work's numerical experiments.}
    \label{fig:loss_gain_spectral}
\end{figure}

\section{Velocity-space dynamics of fusion-product heating in reactive--elastic D--D fusion plasmas}\label{ss:numerical_eg}

We perform a series of numerical examples to illustrate the velocity-space dynamics of a reactive--elastic (fusion) system. Simulations are performed in three velocity dimensions, with $N = 128$ grid points per dimension and $(N_{\hat{q}}, N_{|q|}) = (6, 64)$ quadrature points for the gain term of the irreversible reactive operator. For reactive interactions, we consider only the neutron branch of D--D fusion
\begin{equation}\label{eq:neutron_branch_DD}
\mathrm{D} + \mathrm{D} \rightarrow {}^{3}\mathrm{He}\,(0.82~\mathrm{MeV}) + \mathrm n\,(2.45~\mathrm{MeV}),
\end{equation}
and do not model neutrons explicitly, since they are largely transparent to the
charged species. 


In sizing the domain, we estimate the support of the ${}^{3}\mathrm{He}$ products by adding their center-of-mass birth speed to a Gaussian Doppler shift associated with the reactant center-of-mass velocity, $G$. Thus,
\begin{equation}\label{eq:v_parametrization_boltzR_COM}
    |\bm v_{DD}^{He \, n}|
\lesssim
\sqrt{\frac{2(0.82~\mathrm{MeV})}{m_{He}}}+G .
\end{equation}
For independent Maxwellian deuterium reactants with thermal speed
$v_{\mathrm{th}}=\sqrt{2T_D/m_D}$, the center-of-mass velocity distribution has thermal speed $v_{\mathrm{th}}/\sqrt{2}$. Truncating its support at four times this value gives
\begin{equation}\label{eq:G_bound}
    G = \frac{|\bm v + \bm w|}{2}  \lesssim 6.2 \times 10^6 \mathrm{m/s},
\end{equation}
for $v_{\mathrm{th}}$ corresponding to a $50~\mathrm{keV}$ deuterium population. Taking the reactant support to be $S^- = 4 v_{\mathrm{th}}$, equation \eqref{eq:domain_sizing_energy_cons} gives $R^- = 2S^- =  1.75 \times 10^7\ \mathrm{m/s}$ and \(R^+ = 3.53 \times 10^7\ \mathrm{m/s}\). Finally, substituting \eqref{eq:G_bound} into \eqref{eq:v_parametrization_boltzR_COM}, we get \(|\bm v_{DD}^{He n}| \lesssim 1.4 \times 10^7\ \mathrm{m/s}\), and \(L \approx 2.8 \times 10^7\ \mathrm{m/s}\). 

The first-order IMEX discretization with operator splitting in \eqref{eq:IMEX_LOD_MLB} is used when elastic collisions are treated with the LB operator, whereas a second-order explicit Runge-Kutta scheme is used when they are treated with the Landau collision operator. The numerical results were obtained on an NVIDIA Quadro RTX 6000 GPU (24 GB of memory), leveraging the FFT's parallelism. Typical simulations required $\mathcal{O}(10^{1})$--$\mathcal{O}(10^{2})$ hours of wall-clock time, depending on the number of species and the collisionality strength. Although the number of elastic collision pairs scales quadratically with the number of species, the simulation's runtime does not, as it is dominated by the evaluation of the reactive Boltzmann collision operator, which is computed here for a single reactive channel regardless of the total number of species. In practice, the computation of the reactive Boltzmann operator is more than two orders of magnitude slower than the elastic operator, consistent with the formal computational complexity of the methods used in this work.

\subsection{Fusion product heating in a two-species D--${}^{3}\mathrm{\bf{He}}$ system}\label{ss:num_eg_2species}
\noindent
We consider the two-species (D, ${}^3 \mathrm{He}$) irreversible reactive--elastic system:
\begin{equation}\label{eq:DHe_system}
    \begin{aligned}        
    \partial_tf^{(He)} &= \frac{1}{2} Q_{He}^{R,+}[f^{(D)}, f^{(D)}] + S \cdot Q^E[f^{(He)}, f^{(He)}] + S \cdot Q^E[f^{(He)}, f^{(D)}], \\
    \partial_t f^{(D)} &=  -Q_D^{R,-}[f^{(D)}, f^{(D)}] + S \cdot Q^E[f^{(D)}, f^{(D)}] + S \cdot Q^E[f^{(D)}, f^{(He)}].
    \end{aligned}
\end{equation}
where the reactive operators correspond to the (exoergic) fusion reaction \eqref{eq:neutron_branch_DD}. The parameter $S \leq 1$ is an artificial scaling of the elastic Coulomb operator. It is introduced primarily to make explicit time-stepping of the Landau operator computationally tractable, since for the physical value $S=1$ the Coulomb relaxation time is more than five orders of magnitude shorter than the fusion time scale. 

Reducing $S$ also acts as a diagnostic tool that amplifies the relative strength of the fusion operator, allowing us to observe the velocity-space structure of the reactive source and sink and the formation of the slowing-down distribution, which would otherwise be rapidly suppressed by Coulomb thermalization. We present results for several values of $S$ to assess the sensitivity of the dynamics to the reduced Coulomb collisionality. 

Lastly, note that neglecting the proton branch of D--D fusion effectively halves the deuterium reaction rate. Consequently, the ratio of elastic to reactive operator magnitudes at $S = 1/2$ with the neutron branch only is comparable to that of the physical case, $S = 1$, with both branches included. 

We take a weakly coupled operating point given by
\begin{equation}
n_D = 10^{26}\,\mathrm{m^{-3}}, 
\qquad 
T_D = 10\,\mathrm{keV},
\end{equation}
with initial conditions:
\begin{equation}
f^{(D)}(0,\bm v) 
= n_D
\left(\frac{m_D}{2\pi T_D}\right)^{3/2}
\exp\!\left(-\frac{m_D |\bm v|^2}{2T_D}\right),
\qquad
f^{(He)}(0, \bm v) = 0.
\end{equation}
To quantify deviations from Maxwellian equilibrium during the evolution, we use a scalar diagnostic introduced in \cite{DattaShumlak2023}:
\begin{equation}
\chi_i(t) = \frac{\|f^{(i)}(\cdot,t) - M_f^{(i)}(\cdot,t)\|_F}{\|M_f^{(i)}(\cdot,t)\|_F},
\end{equation}
where $M_f^{(i)}$ denotes the equivalent Maxwellian with the same instantaneous density, mean velocity, and temperature as $f^{(i)}$, and $\|\cdot\|_F$ is the Frobenius norm over the velocity grid. 

Using equation \eqref{fureact}, we also evaluate the fusion reactivity $\langle \sigma \bm v \rangle$ of $f^{(i)}$ and compare it with the (Maxwellian) reactivity $\langle \sigma  \bv \rangle_M$ of $M_f^{(i)}$. $\chi$ may be viewed as a uniform measure of deviation from the equivalent Maxwellian, while the reactivity ratio, $\langle \sigma \bv \rangle / \langle \sigma \bv \rangle_M$, measures this deviation weighted by the fusion cross-section.

\begin{figure}[H]
    \centering
    \subcaptionbox{\label{fig:S5en3_combined_a}}[0.49\textwidth]{
    \includegraphics{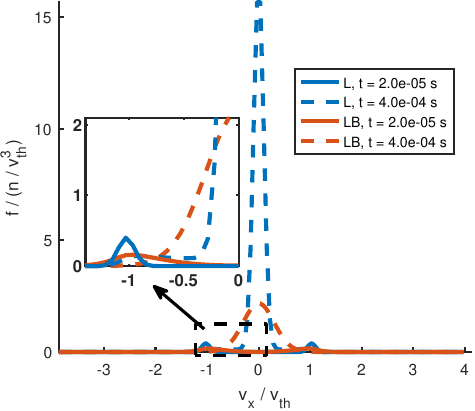}}
    \subcaptionbox{\label{fig:S5en3_combined_b}}[0.49\textwidth]{
    \includegraphics{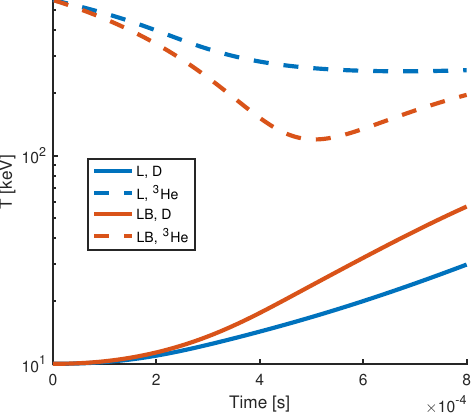}}
    \caption{Fusion product heating in a burning $\mathrm D-{}^3 \mathrm{He}$ plasma with $S = 2.5 \times 10^{-3}$. Elastic collisions are treated with the Landau (L) and Lenard--Bernstein (LB) operators; a time-step of $\Delta t = 2\times10^{-7}$ s is used in both cases. (a) Normalized ${}^3$He velocity distribution function, $(f^{(He)} / (n_{He}(t) / v_{\mathrm{th}}^3))$. The solution illustrates the expected formation and relaxation of the fusion-product distribution, forming a ``slowing down'' distribution with a thermal (``ash'') component and a heavy tail. (b) Temperature evolution showing bulk heating of the fuel due to elastic scattering with energetic fusion products.}
    \label{fig:S5en3_combined}
\end{figure}

\begin{figure}[H]
    \centering
    \includegraphics{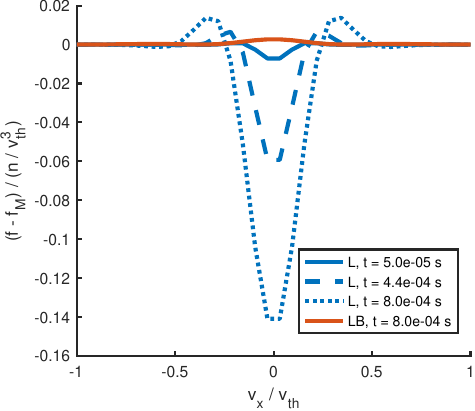}
    \caption{Normalized deviation of the deuterium distribution from its equivalent Maxwellian, $(f^{(D)} - M_f^{(D)}) \, / \, (n_D(t) / v_{\mathrm{th}}^3)$, with $S = 2.5 \times 10^{-3}$. The depletion is concentrated in the velocity range where the fusion rate is maximized, determined by the product of the fusion cross-section and the reactant densities. The high density of slow particles in a near-Maxwellian population leads to frequent interactions with energetic particles, enhancing their depletion. Note that the peak magnitude of this depletion is three orders of magnitude smaller than the peak of the normalized distribution functions in Fig.~\ref{fig:S5en3_combined}, reflecting the reactants' small deviation from the equivalent Maxwellian distribution. An accumulation of mass, relative to the Maxwellian equilibrium, occurs at the energetic tails of the deuterium distribution function. The LB solution, shown at the simulation end time, is not able to capture the aforementioned velocity-dependent deviations from Maxwellian equilibrium.}
    \label{fig:fD_minus_fDM}
\end{figure}

The numerical results demonstrate that the LB operator cannot resolve the suprathermal features observed in the Landau solution, as shown in Figs.~\ref{fig:S5en3_combined_a} and \ref{fig:fD_minus_fDM}. The former damps the suprathermal tail too rapidly. This stems from the formulation of the operator, which tunes a velocity-independent collision frequency and mixture Maxwellian coefficients to match conservation properties, an $\mathcal{H}$-theorem, and the near-equilibrium momentum and temperature relaxation rates of the Boltzmann--Coulomb operator, but does not involve information about higher-order moments of the solution, let alone its detailed velocity-space structure. The velocity-independent collision frequency cannot capture the reduced collisionality of fast particles, necessarily overestimating their depletion and failing to resolve the correct velocity distribution function. 

Figure~\ref{fig:S5en3_combined_b} shows that the LB model also overestimates the inter-species temperature relaxation rate, consistent with our observation in Sec.~\ref{ss:elastic_relaxation_rate_matching}. This is because its velocity-independent collision frequency is tuned to match the moment relaxation rates of the Boltzmann--Coulomb operator in a near-Maxwellian regime, as outlined in \cite{Morse_PhysFluids_1963}. For these rates to hold, it is required not only that the initial distribution functions are close to Maxwellian, but also that they remain so throughout the inter-species relaxation process; this could be satisfied, for instance, if Grad's epochal relaxation hypothesis \cite{Grad1960} holds: intra-species equilibration is much faster than inter-species equipartition. However, this is not satisfied in our application of interest, where energetic ${}^3 \mathrm{He}$ are born far from equilibrium and the species density and temperature ratios are large. The low ${}^3 \mathrm{He}$ density means that it relaxes primarily through inter-species collisions, remaining far-from-Maxwellian throughout the relaxation process and violating the assumptions of the LB model. This operator is therefore not suitable for self-consistent non-Maxwellian reactivity calculations. Thus, we focus our discussion mostly on the Landau--reactive-Boltzmann simulations.

\begin{figure}[H]
    \centering
    \subcaptionbox{$\chi$ Parameter}[0.5\textwidth]
    {\includegraphics{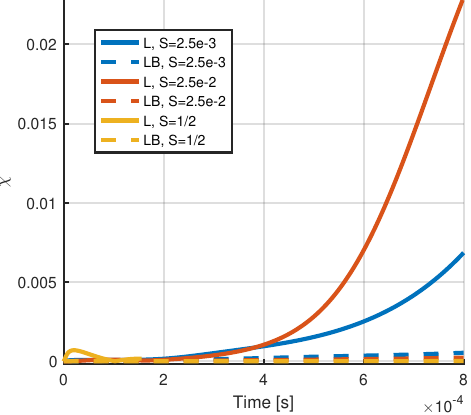}}
    \subcaptionbox{Reactivity Ratio}[0.5\textwidth]
    {\includegraphics{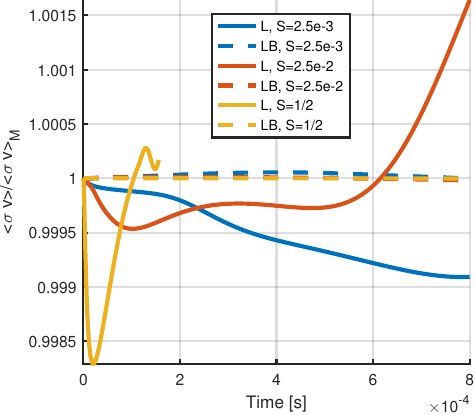}}
    \caption{Deviation of the deuterium velocity distribution function from its equivalent Maxwellian at varying collisionality, S, as measured by (a) the $\chi$ parameter and (b) the reactivity ratio. The strongest collisionality ($S = 1/2$) exhibits more pronounced non-Maxwellian features at early times than the weaker cases, but these are rapidly damped by Coulomb thermalization. These results are shown only at early times due to their high computational cost. By both scalar metrics, the early-time deviation from Maxwellian remains below $0.2\%$ across all collisionalities. At later times, simulations with artificially reduced collisionality reach $\chi \approx 2\%$ and correspondingly large reactivity enhancement, but this corresponds to an unphysically high deuterium temperature of $30$ to $50~\mathrm{keV}$, resulting from the absence of spatial transport to offset heating by fusion products.}
    
    \label{fig:chi_react_ratio}
\end{figure}

There is no significant deviation of the reactant distribution from its equivalent Maxwellian, with a reactivity enhancement of less than $0.05\%$ for the $S = 1/2$ case, confirming the validity of the Maxwellian reactant approximation in the simulated regime. Fig.~\ref{fig:S5en3_combined_a} illustrates the velocity-space dynamics of the system, with the solution normalized by the time-varying species density divided by a characteristic velocity-space volume associated with the thermal speed $v_{\mathrm{th}} = L/4$. Fig.~\ref{fig:S5en3_combined_b} displays the expected bulk heating of the plasma due to fusion reactions. Fig.~\ref{fig:fD_minus_fDM} shows that, relative to its equivalent Maxwellian, the deuterium distribution develops a slight enhancement in the high-velocity tail due to Coulomb collisions with energetic fusion products. All simulations exhibit an initial depletion of the energetic deuterium tail due to fusion, resulting in a transient reduction in reactivity relative to the Maxwellian value; as the ${}^3\mathrm{He}$ population accumulates, Coulomb collisions progressively replenish the tail, leading to a modest reactivity increase (Fig.~\ref{fig:chi_react_ratio}). The onset of this replenishment occurs on a timescale comparable with the deuterium--helium-3 energy equilibration time.

By the end of the simulations with $S = 2.5\times 10^{-3}$ and $S = 2.5\times 10^{-2}$, the deuterium temperature exceeds $30~\mathrm{keV}$, reducing the rate of energy transfer from the energetic ${}^3\mathrm{He}$ population to the background fuel. As a result, the initial cooling of ${}^3\mathrm{He}$ is arrested, and its temperature subsequently increases due to continued injection of energetic fusion products. 

Oscillations in $\chi$ of order $10^{-5}$ are comparable to the numerical error in elastic-only simulations, where we observed $\chi=\mathcal{O}(10^{-5})$ near the final time, rather than its exact equilibrium value of zero. These oscillations reflect the accumulation of small conservation errors by the fast Fourier spectral method at finite velocity-space resolution. By contrast, the larger temporal trends in $\chi$ and the systematic initial decrease and subsequent increase in the reactivity ratio are attributable to the competing effects of fusion-driven tail depletion and collisional replenishment described above.

\subsection{Fusion product heating in a three-species $\mathrm D-{}^{3}\mathrm{He}-\mathrm e$ system}\label{ss:3species_num_eg}

We consider the same domain sizing, velocity grid, density, and D--D reactive channel as in Sec.~\ref{ss:num_eg_2species}, this time for a three-species (D, ${}^3$He, e) system
\begin{equation}\label{eq:DHee_system}
    \begin{aligned}     
        \partial_tf^{(He)} &= \frac{1}{2} Q_{He}^{R,+}[f^{(D)}, f^{(D)}] + S \cdot Q^E[f^{(He)}, f^{(He)}] + S \cdot Q^E[f^{(He)}, f^{(D)}] + S \cdot Q^E[f^{(He)}, f^{(e)}], \\
        \partial_t f^{(D)} &= -Q_D^{R,-}[f^{(D)}, f^{(D)}] + S \cdot Q^E[f^{(D)}, f^{(D)}] + S \cdot Q^E[f^{(D)}, f^{(He)}] + S \cdot Q^E[f^{(D)}, f^{(e)}], \\
        \partial_t f^{(e)}&= S \cdot Q^E[f^{(e)}, f^{(e)}] + S \cdot Q^E[f^{(e)}, f^{(D)}] + S \cdot Q^E[f^{(e)}, f^{(He)}],
    \end{aligned}
\end{equation}
where $e$ denotes the electron with an artificial deuteron-to-electron mass ratio of ten. The electrons do not participate in reactions; i.e., $Q_e^{R,+} = Q_e^{R,-} = 0$.

\begin{figure}[H]
    \centering
    \includegraphics{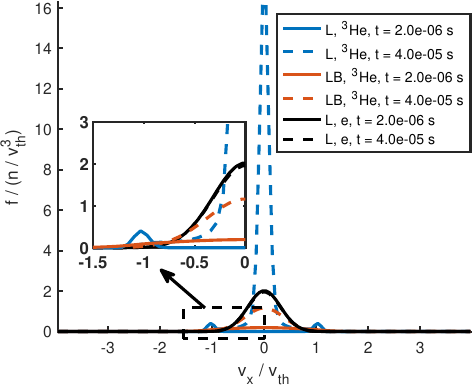}
    \caption{Normalized helium-3 (${}^3 \mathrm{He}$) and electron (e) velocity distribution functions central slice with $S = 2.5 \times 10^{-3}$ at different times. The ${}^3 \mathrm{He}$ solutions are given for both a Landau (L) and Lenard--Bernstein (LB) treatment of elastic collisions. This central velocity slice illustrates the formation and relaxation of the fusion-product distribution, exhibiting a ``slowing down'' profile with a thermal (``ash'') component and a suprathermal tail. We use $\Delta t = 4 \times 10^{-10}$ s for the Landau solution and $\Delta t = 4 \times 10^{-8}$ s for the Lenard--Bernstein solution.}
    \label{fig:S1_3sp_fslice}
\end{figure}

\begin{figure}[H]
    \centering
    \includegraphics{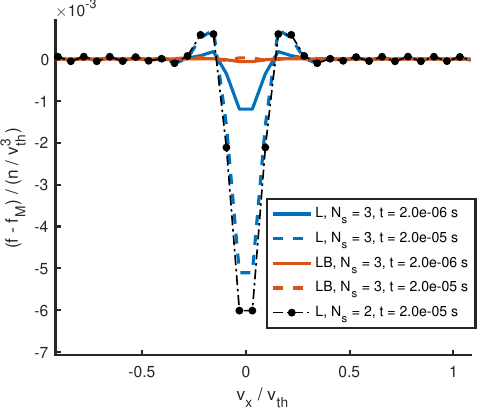}
    \caption{Normalized deviation of the deuterium distribution from its corresponding Maxwellian, $(f^{(D)} - M_f^{(D)}) \, / \, (n_D(t) / v_{\mathrm{th}}^3)$, with $S = 2.5 \times 10^{-3}$. The depletion structure in the three-species ($N_s = 3$) system is similar to that of the two-species ($N_s = 2$) results, shown for comparison. The peak depletion magnitude remains small compared to the normalized distribution functions in Fig.~\ref{fig:S1_3sp_fslice}.}
    \label{fig:S1_3sp_fD_dev}
\end{figure}

We note that the reduced mass ratio preserves the qualitative structure of the energy transfer cascade, in which fusion-product energy is transferred to the electrons more rapidly than to the deuterium bulk, and electron--deuterium equipartition occurs on a longer timescale. However, it does not capture the correct magnitude of the separation of timescales. For the physical mass ratio, the much lighter electrons equilibrate significantly faster than the ions, which can modify the balance between energy transfer to the deuterium population and its depletion due to fusion. Nevertheless, this scaling provides a tractable first step toward capturing the essential physics of the coupled reactive and collisional dynamics in a three-species system with disparate mass ratios. We again consider a weakly coupled operating point with initial deuterium and electron densities
$n_D(0)=n_e(0)=10^{26}\,\mathrm{m^{-3}}$ and temperatures
\begin{equation}
T_D = 10\,\mathrm{keV},
\qquad
T_e = 10\,\mathrm{keV}.
\end{equation}
The initial conditions are
\begin{equation}
\begin{aligned}
f^{(D)}(0,\bm v) 
&= n_D
\left(\frac{m_D}{2\pi T_D}\right)^{3/2}
\exp\!\left(-\frac{m_D |\bm v|^2}{2 T_D}\right), \\
f^{(He)}(0,\bm v) &= 0, \\
f^{(e)}(0,\bm v) 
&= n_e
\left(\frac{m_e}{2\pi T_e}\right)^{3/2}
\exp\!\left(-\frac{m_e |\bm v|^2}{2 T_e}\right).
\end{aligned}
\end{equation}
As in the two-species case, we observe no significant deviation from Maxwellian behavior in the reactants. The inclusion of electrons substantially increases the temporal stiffness of the system. As a result, only early-time results are provided, which primarily serve as verification of consistency with the early-time physical mechanisms identified in the two-species case. Figures~\ref{fig:S1_3sp_fslice}--~\ref{fig:S1_3sp_fD_dev} show that the addition of electrons does not qualitatively alter the velocity-space dynamics: the ${}^3 \mathrm{He}$ distribution exhibits the expected slowing-down profile with a thermal (``ash'') component and a suprathermal tail, while $(f^{(D)} - M_f^{(D)})$ retains a depletion structure similar to the two-species case. Further, we observe no appreciable enhancement in reactivity or sustained suprathermal reactant population. This agrees qualitatively with recent kinetic simulations of burning D--T plasmas~\cite{vandeWetering2025PICNIC_NES_ICF}, where small-angle (Landau-type) elastic collisions produce a small suprathermal ion population compared to results with large-angle elastic scattering (Boltzmann-Coulomb-type).

\begin{figure}[H]
    \centering
    \subcaptionbox{}[0.50 \textwidth]
    {\includegraphics{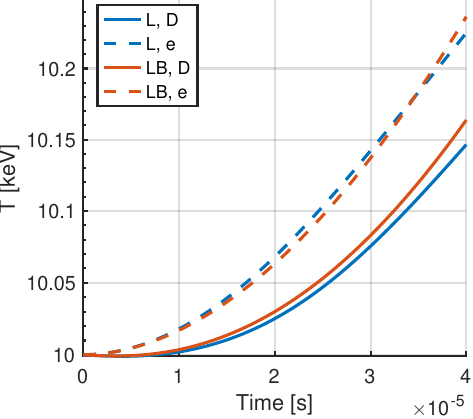}}
    \par\medskip
    \subcaptionbox{}[0.50 \textwidth]
    {\includegraphics{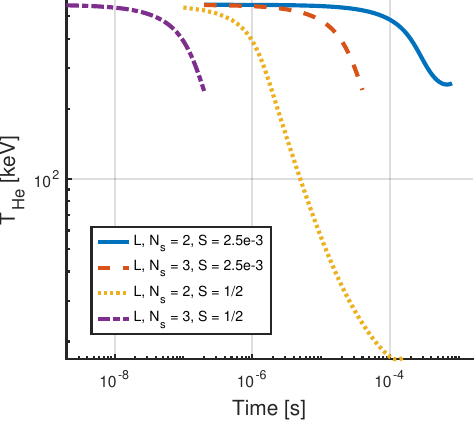}}
    \caption{Temperature evolution in the three-species ($N_s = 3$) system. (a) Deuterium and electron temperature evolution at $S = 2.5 \times 10^{-3}$, showing the bulk fuel heating driven by energy deposition from fusion products. The lighter electrons heat more rapidly than the deuterium. (b) ${}^3\mathrm{He}$ temperature evolution, with two-species ($N_s = 2$) results shown for comparison. Each curve begins at the first stored positive time, since $T_{\mathrm{He}}$ is undefined at $t=0$ when $n_{\mathrm{He}}=0$. The presence of electrons in the three-species simulations accelerates the cooling of fusion products.}
    \label{fig:S5en3_3sp_kT_combined}
\end{figure}

\begin{figure}[H]
    \centering
    \subcaptionbox{$\chi$ Parameter}[0.5\textwidth]
    {\includegraphics{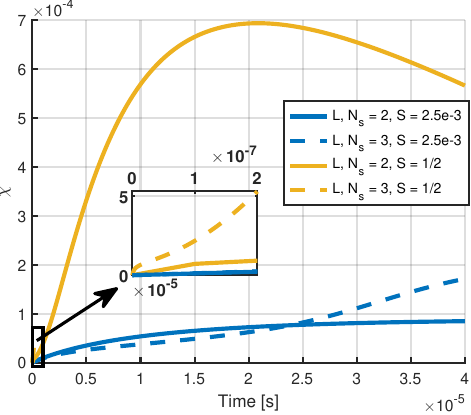}}
    \subcaptionbox{Reactivity Ratio}[0.5\textwidth]
    {\includegraphics{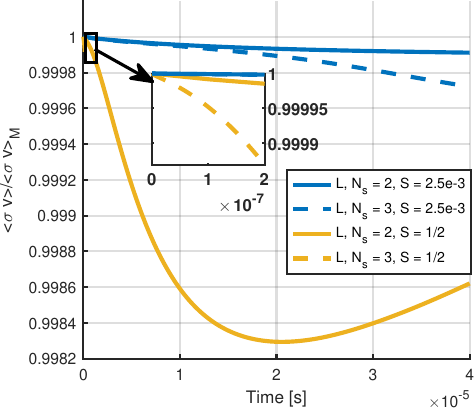}}
    \caption{Deviation of the deuterium distribution function from its equivalent Maxwellian in two-species ($N_s = 2$) and three-species ($N_s = 3$) systems for varying collisionality $S$, as measured by (a) the $\chi$ parameter and (b) the reactivity ratio. Temporal stiffness restricts the $S=1/2$ three-species simulation to early times, over which deviations from Maxwellian behavior remain small. The initial reduction in the reactivity ratio is associated with preferential fusion depletion of the energetic deuterium tail. This reduction is slightly faster in the three-species system than in the two-species system.}
    \label{fig:S1_3sp_chi_react}
\end{figure}

The temperature evolution and scalar diagnostics are consistent with the expected energy partitioning in the three-species system. As shown in Fig.~\ref{fig:S5en3_3sp_kT_combined}, the electrons heat more rapidly than the deuterium, and their presence increases the rate at which the ${}^3\mathrm{He}$ population is cooled, as fusion-product energy is shared between more species. Correspondingly, Fig.~\ref{fig:S1_3sp_chi_react} shows that both the $\chi$ parameter and the reactivity ratio initially decrease due to depletion of the energetic deuterium tail prior to collisional replenishment, with deviations comparable in magnitude to the two-species case. The inclusion of electrons slightly accelerates this deviation, and no reactivity enhancement is observed over the accessible simulation timescale.

\section{Conclusion}

We have presented a fully kinetic, multi-species framework for the self-consistent evolution of reacting, collisional plasmas with arbitrary reactant distribution functions. The numerical results establish, for the first time within a continuum kinetic setting, that the Maxwellian reactant approximation is justified in the weakly coupled, spatially homogeneous, early-time D--D fusion regimes considered. Throughout the early and intermediate stages of the simulations, the reactants' relative Frobenius-norm deviation from Maxwellian remains $\mathcal{O}(10^{-3})$, with larger deviations arising only at late times in cases with artificially reduced collisionality. Likewise, the fusion reactivity ratio deviates from unity by $\mathcal{O}(10^{-3})$ or less, with an initial reduction followed by a small enhancement of the reactivity relative to its Maxwellian equivalent. The LB operator fails to capture the velocity-space structure of the non-Maxwellian reactant distribution and, consequently, the correct profiles of the reactivity ratio and relative Frobenius norm deviation $\chi$. This limitation and the faster temperature relaxation rates observed in Sec.~\ref{ss:elastic_relaxation_rate_matching} and \ref{ss:numerical_eg} are consistent with the model's velocity-independent effective collision frequency, which cannot capture the reduced collisionality of fast particles, as well as its tuning to near-Maxwellian moment relaxation rates. Together, these effects limit the operator's utility for non-thermal reactivity calculations.

These results demonstrate that no appreciable suprathermal deuterium population develops under the conditions studied. This can be understood in terms of the underlying energy transfer mechanisms. In the present model, fusion heating is treated self-consistently, in contrast to prior studies with prescribed ion distributions or possibly unphysically strong heating sources. Moreover, D--D fusion products are significantly less energetic than the alpha particles produced in D--T reactions, leading to presumably weaker suprathermal effects. Finally, only cumulative small-angle Coulomb collisions, modeled via the Landau and LB operators, were included. Rare large-angle Coulomb collisions and nuclear elastic scattering were neglected, though they have been shown to enhance suprathermal populations in moderately coupled D--T plasmas~\cite{vandeWetering2025PICNIC_NES_ICF}.

 The fast spectral method for the reactive Boltzmann collision operator provides a computationally efficient way to evaluate reactivity integrals and reactive sources for arbitrary distribution functions. Together with the elastic collision terms, the framework provides a pathway to systematically assess departures from the Maxwellian reactant assumption in regimes where higher fusion product energies, stronger heating, or large-angle elastic scattering mechanisms are present. Such regimes are relevant to D--T fusion plasmas, external heating mechanisms such as neutral beam injection, and moderately coupled plasmas. Extensions to spatially inhomogeneous systems would allow fusion-product heating to be balanced by transport, rather than increasing monotonically in time as in the present spatially homogeneous setting. They would also enable the inclusion of magnetic fields, which can drive anisotropic non-Maxwellian effects. The reactive--elastic Boltzmann model naturally admits such generalizations, but we defer this to future work due to its computational cost. While the fast spectral method mitigates the computational cost of integro-differential collision operators in high-dimensional velocity space, the D--T burning plasma regime remains more challenging than the D--D cases considered in this work, due to higher fusion product energies and additional species, amplifying multiscale structure in velocity space and temporal stiffness. An efficient continuum kinetic implementation may leverage low-rank tensor approximation \cite{Einkemmer2025LowRankKinetic} of the collision terms, penalization-type methods \cite{FJ_JCP_2010} for IMEX time-integration of Boltzmann-type operators, and non-uniform velocity meshes to address these difficulties.

 \section*{Acknowledgments}

 The information, data, or work presented herein is based in part upon work supported by the National Science Foundation under Grant No. IIS-2433957 and by the Air Force Office of Scientific Research under Award No. FA9550-26-1-B161. This work was also partially supported by DOE grant DE-SC0023164, NSF grant DMS-2409858, and DoD MURI grant FA9550-24-1-0254.

 \section*{Data Availability}

The data that support the findings of this study are available from the corresponding author upon reasonable request.



\printbibliography




\end{document}